\documentclass[aps,prl,twocolumn,preprintnumbers,amsmath,amssymb,superscriptaddress,floatfix]{revtex4}
\usepackage{graphicx}
\usepackage{dcolumn}
\usepackage{bm}
\usepackage{amsfonts}
\usepackage{amsmath}
\usepackage{bm}% bold math
\usepackage{amsmath}% needed for subequations

\begin{document}
\title{Pseudospin magnetism in graphene}
\author{Hongki Min}
\email{hongki@physics.utexas.edu}
\affiliation{Department of Physics, The University of Texas at Austin, Austin, Texas 78712, USA}
\author{Giovanni Borghi}
\affiliation{NEST-CNR-INFM and Scuola Normale Superiore, I-56126 Pisa, Italy}
\author{Marco Polini}
\affiliation{NEST-CNR-INFM and Scuola Normale Superiore, I-56126 Pisa, Italy}
\author{A.H. MacDonald}
\affiliation{Department of Physics, The University of Texas at Austin, Austin, Texas 78712, USA}

\date{\today}
%\date{Dated: July 10, 2007}

\begin{abstract}
We predict that neutral graphene bilayers are pseudospin magnets in which the
charge density-contribution from each valley and spin spontaneously shifts to one of
the two layers.  The band structure of this system is characterized by a
momentum-space vortex which is responsible for unusual competition between band and
kinetic energies leading to symmetry breaking in the vortex core.  We discuss the possibility of realizing
a pseudospin version of ferromagnetic metal spintronics in graphene bilayers based on
hysteresis associated with this broken symmetry.
\end{abstract}

\maketitle
\normalsize

\noindent {\it Introduction}---The ground state of an interacting electron system
flows from subtle compromises between band and interaction energy minimization.  Because of the Pauli blocking
effects which underlie Fermi liquid theory however, the consequences of interactions are
normally only quantitative~\cite{Giuliani_and_Vignale} unless symmetries are broken.
Recent progress~\cite{novoselov2004} in the isolation of single and double sheets of carbon atoms (graphene sheets) has presented
researchers with a new type of interacting electron system whose properties are now being actively explored~\cite{novoselov2005,zhang2005,Geim_MacDonald}, both theoretically and experimentally.  In this Letter we argue that band energy minimization is exceptionally
frustrating to interactions in graphene bilayers, and predict that broken symmetry
states in which charge shifts spontaneously from one layer to the
other occur as a consequence.

\begin{figure}[h]
\includegraphics[width=1.0\linewidth]{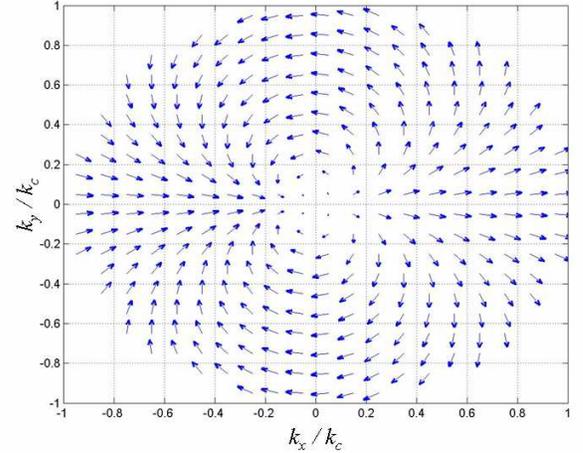}
\caption{Pseudospin orientation in a graphene bilayer broken symmetry state.
In this figure the arrows represent both the magnitude and the direction of the $\hat{x}$-$\hat{y}$
projection of the pseudospin orientation $\hat{n}$ as obtained from a mean-field-theory calculation
for a neutral, unbiased bilayer with coupling constant $\alpha=1$.
%{\bf Hongki:  Insert value here.}
The arrows are shorter in the core of the momentum space vortex because the pseudospins in the core have
rotated spontaneously toward the $\hat{z}$ or $-\hat{z}$ direction.
%{\bf Hongki: Is this really a color figure as the caption suggests - if not remove `color online'.}
\label{fig:pseudospin}}
\end{figure}

Graphene bilayers with Bernal stacking have one low-energy site per unit cell in each layer.
When the layer degree of freedom is described as a pseudospin, the continuum limit of the $\pi$-orbital
band Hamiltonian corresponds~\cite{novoselov2006,mccann2006} to a pseudospin field
${\bm B}_{\rm band} = [\hbar^2 k^2/(2m^\star)] \, (\cos(2 \phi_{\bm k}),\sin(2 \phi_{\bm k}), 0)$,
where $\phi_{\bm k}= \arctan{(k_y/k_x)}$, $m^\star=\gamma_1/(2 v^2_{\rm F})$,
$\gamma_1$ is the interlayer tunneling amplitude, and $v_{\rm F}$
is the electron velocity at the Fermi energy in an isolated neutral graphene sheet.
When interactions are neglected the ground state of a neutral bilayer has a full valence
band of pseudospinors aligned at each ${\bm k}$ with
this pseudospin field, forming the momentum-space vortex.
%illustrated in Fig.[~\ref{fig:pseudospin}].
%Hongki: This figure is obtained with interactions
The vortex exacts a large 
interaction energy penalty because of its rapid pseudospin-orientation variation.
We propose that, like its real-space counterpart~\cite{vortex}, the momentum-space vortex
sidesteps this energy cost by forming a vortex core in which the pseudospin
orientation is out of plane in either the $\hat{z}$ or $-\hat{z}$ direction,
as illustrated in Fig.~\ref{fig:pseudospin}.
The momentum-space vortex state is nonuniform in momentum space, but in real space
transfers charge uniformly between layers. Our paper starts by describing a technical calculation that supports and elaborates on our prediction
and then discusses anticipated properties of this state.

\noindent {\it Chiral two-dimensional electron system Hartree-Fock Theory}---It is instructive to consider a class of chiral two-dimensional electron system (C2DES) models which includes the continuum limits of single-layer and bilayer graphene sheets as special cases. These C2DES models have band Hamiltonians,
\begin{widetext}
\begin{equation}
{\hat {\cal H}}_{\rm band} = -\sum_{{\bm k}, \sigma', \sigma} \;
{\hat c}^{\dagger}_{{\bm k}, \sigma'}
\left\{\varepsilon_0(k_{\rm c})\left({k\over k_{\rm c}}\right)^{J}\left[\cos(J \phi_{\bm k}) \, \tau^{x}_{\sigma',\sigma}
+ \sin(J \phi_{\bm k}) \, \tau^{y}_{\sigma',\sigma}\right] + {V_{\rm g}\over 2}~\tau^{z}_{\sigma',\sigma} \right\}
{\hat c}_{{\bm k}, \sigma}\,,
\label{eq:band}
\end{equation}
\end{widetext}
where $\sigma,\sigma'$ are pseudospin labels, $J$ is the chirality index, $\tau^{a}$ is a Pauli matrix,
$k_{\rm c}$ is the model's ultraviolet momentum cutoff,
% whose role we discuss below,
$\varepsilon_0(k_{\rm c})$ is the energy scale of the band Hamiltonian, and a sum over valley and
spin components is implicit.
In Eq.~(\ref{eq:band}), $V_{\rm g}$ is an external potential term which couples to the
pseudospin magnet order parameter and corresponds in the case of bilayer
graphene to an external potential difference between the layers.
For single-layer graphene $J=1$ and $\varepsilon_0(k_{\rm c})=\hbar v_{\rm F} k_{\rm c}$,
while for bilayer graphene $J=2$ and $\varepsilon_0(k_{\rm c})=\hbar^2 k^2_{\rm c}/(2m^\star)$.
The dimensionless coupling constant of these C2DESs, which measures the interaction
strength,  can be defined as $\alpha=(e^2 k_{\rm c}/\epsilon)/\varepsilon_0(k_{\rm c})$
where $\epsilon$ is the effective dielectric function due to screening
external to the $\pi$ electron system. In the case of a single graphene layer $\alpha_{\rm mono}=e^2/(\epsilon v_{\rm F}\hbar)$,
while in the bilayer case, $\alpha_{\rm bi}=2e^2/(\epsilon v_{\rm c}\hbar)$, where $v_{\rm c}=\hbar k_{\rm c}/m^\star$.
If we choose~\cite{min2007} $\hbar k_{\rm c} = \sqrt{2 m^\star \gamma_1}$ for the bilayers,
we have $\alpha_{\rm bi}=\alpha_{\rm mono}$. Typically $\epsilon \sim 2.5$ which implies a dimensionless coupling constant
$\alpha\sim 1$.  We use $\varepsilon_0(k_{\rm c})$ and $k^{-1}_{\rm c}$ as energy and length units in the rest of this paper.

The C2DES Hartree-Fock Hamiltonian can be written (in dimensionless units) in the following physically transparent form:
\begin{equation}
{\hat {\cal H}}_{\rm HF} = -\sum_{{\bm k}, i, \sigma', \sigma}
{\hat c}^{\dagger}_{{\bm k}, i, \sigma'}{\cal B}^{(i)}_{\sigma',\sigma}({\bm k}){\hat c}_{{\bm k}, i, \sigma}\,,
\label{eq:HF}
\end{equation}
where  ${\cal B}^{(i)}_{\sigma',\sigma}=B^{(i)}_0({\bm k})\delta_{\sigma',\sigma}+ {\bm B}^{(i)}(\bm{k}) \cdot {\bm \tau}_{\sigma', \sigma}$,
\begin{equation}\label{eq:B0}
B^{(i)}_0({\bm k})=\alpha\int_{|{\bm k}'|<1}\frac{d^2{\bm k}'}{2\pi}\frac{1}{|{\bm k}-{\bm k}'|}
\frac{f^{(i)}_{\rm sum}({\bm k}')}{2}\,,
\end{equation}
and the pseudospin field ${\bm B}^{(i)}({\bm k})$ has band and interaction contributions,
\begin{widetext}
\begin{equation}
\label{eq:Bx}
B^{(i)}_x({\bm k}) = k^{J} \cos(J \phi_{\bm k})
+\alpha \int_{|{\bm k}'|<1} \frac{d^2{\bm k'}}{2\pi} {e^{-|{\bm k}-{\bm k'}|{\bar d}}\over |{\bm k}-{\bm k'}|} {f^{(i)}_{\rm diff}({\bm k'}) \over 2} \; n^{(i)}_x({\bm k'})\,,
\end{equation}
\begin{equation}
\label{eq:By}
B^{(i)}_y({\bm k}) = k^{J} \sin(J \phi_{\bm k})
+\alpha \int_{|{\bm k}'|<1} \frac{d^2{\bm k'}}{2\pi} {e^{-|{\bm k}-{\bm k'}|{\bar d}}\over |{\bm k}-{\bm k'}|} {f^{(i)}_{\rm diff}({\bm k'}) \over 2} \; n^{(i)}_y({\bm k'})\,,
\end{equation}
\begin{equation}
\label{eq:BzA}
B^{(i)}_z({\bm k}) = \frac{{\bar V}_g}{2}
+\alpha\sum_{j} \int_{|{\bm k}'|<1} \frac{d^2{\bm k'}}{2\pi}
\left( {1\over |{\bm k}-{\bm k'}|}\, \delta_{i,j} -  {\bar d} \right)
{f^{(j)}_{\rm diff}({\bm k'}) \over 2} \; n^{(j)}_z({\bm k'})\,.
\end{equation}
\end{widetext}
Here $i,j$ label the four valley and spin components of graphene's $J=1$ and $J=2$ C2DESs,
${\bm n}^{(i)}({\bm k})$ is the direction of ${\bm B}^{(i)}({\bm k})$, $f^{(i)}_{\rm sum}({\bm k'})$ [$f^{(i)}_{\rm diff}({\bm k'})$] is the sum of (difference between)
low- and high-energy occupation numbers, ${\bar V}_{\rm g}=V_{\rm g}/\varepsilon_0(k_{\rm c})$ is the gate potential in units of $\varepsilon_0(k_{\rm c})$,
${\bar d}=k_c d$ is the distance between layers in units of $k_c^{-1}$ in the bilayer case,
and ${\bar d}=0$ in the monolayer case.
The term proportional to ${\bar d}$ on the right-hand side of Eq.~(\ref{eq:BzA}) is the Hartree potential which opposes charge transfer between layers
in the bilayer case.

Local minima of the Hartree-Fock energy functional solve Eqs.~(\ref{eq:Bx})-(\ref{eq:BzA})
self-consistently.  Our focus here is on the broken symmetry
momentum-space vortex solutions
in which pseudospins near ${\bm k}=0$ tilt away
from their band Hamiltonian $\hat{x}$-$\hat{y}$ plane orientations toward the $\pm \hat{z}$ direction, {\em i.e.},
${\bm n}^{(i)}(\bm{k})= (n^{(i)}_{\perp}(k) \cos(J \phi_{\bm k}), n^{(i)}_{\perp}(k) \sin(J \phi_{\bm k}),n^{(i)}_z(k))$ with
$[n^{(i)}_{\perp}(k)]^2+[n^{(i)}_z(k)]^2=1$.  Pseudospin polarization in the $\hat{z}$-direction corresponds to charge
transfer between layers.  This {\em ansatz} yields effective magnetic
fields whose $\hat{x}$-$\hat{y}$ plane projections are parallel to the band Hamiltonian effective field.
We find that ${\bm B}^{(i)}(\bm{k})= (B^{(i)}_{\perp}(k) \cos(J \phi_{\bm k}), B^{(i)}_{\perp}(k) \sin(J \phi_{\bm k}),B^{(i)}_z(k))$ with
\begin{equation}
B^{(i)}_{\perp}(k) = k^{J} +  \alpha \int_{0}^{1} dk' F_{\perp}(k,k') \;f^{(i)}_{\rm diff}(k') \; n^{(i)}_{\perp}(k'),
\label{eq:Bperp}
\end{equation}
\begin{widetext}
\begin{equation}
B^{(i)}_z(k) = \frac{{\bar V}_g}{2} + \alpha \sum_{j} \int_{0}^{1} dk'\left(F_{z}(k,k')\delta_{i,j} - {1\over 2} k' {\bar d}\right) \;f^{(j)}_{\rm diff}(k') \; n^{(j)}_{z}(k')\,,
\label{eq:Bz}
\end{equation}
where the exchange kernels in Eqs.~(\ref{eq:Bperp}) and~(\ref{eq:Bz}) are given by
\end{widetext}
%\begin{equation}\label{eq:kernels}
%\left\{
%\begin{array}{l}
%{\displaystyle F_{\perp}(k,k') = k'\int_{0}^{\pi} \; {d\phi \over 2\pi} \; %{e^{-q {\bar d}}\over  q} \cos(J\phi)}\vspace{0.1 cm}\\
%{\displaystyle F_{z}(k,k') = k'\int_{0}^{\pi} \; {d \phi \over 2\pi} \;  %{1\over q}}
%\end{array}
%\right.
%\end{equation}
\begin{eqnarray}
\label{eq:kernels}
F_{\perp}(k,k') &=& k'\int_{0}^{\pi} \; {d\phi \over 2\pi} \; {e^{-q {\bar d}}\over  q} \cos(J\phi), \nonumber \\
F_{z}(k,k') &=& k'\int_{0}^{\pi} \; {d \phi \over 2\pi} \;  {1\over q},
\end{eqnarray}
with $q=q(k,k',\phi)\equiv \sqrt{k^2+k'^2-2k k'\cos(\phi)}$. The pseudospin-chirality induced frustration is represented by the factor $\cos(J\phi)$ in
the first line of Eq.~(\ref{eq:kernels}) which makes $F_{\perp}$ much smaller than $F_{z}$.

\begin{figure}
\begin{center}
\includegraphics[width=1.0\linewidth]{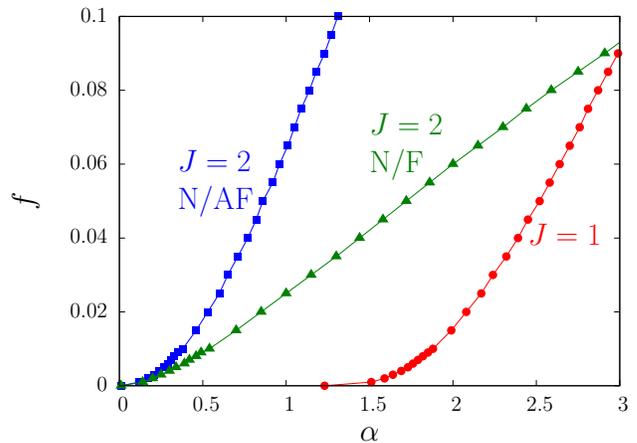}
\caption{(Color online) Phase diagram of C2DES's with $J=2$ and $J=1$.  For the $J=2$ bilayer case we have taken ${\bar d}=0.2$.
Pseudospin magnetism occurs at strong coupling $\alpha$ and weak doping $f$.
%($1+f \equiv \sum_{{\bm k}, i} f^{(i)}_{\rm sum}({\bm k}) /{\cal N}$ where ${\cal N}=\sum_{{\bm k}, i}1$.)
($1+f=n_{\uparrow}+n_{\downarrow}$ where the pseudospin density $n_{\sigma}= \sum_{{\bm k}, i}\langle {\hat c}^\dagger_{{\bm k}, i, \sigma} {\hat c}_{{\bm k}, i, \sigma}\rangle/{\cal N}$ and ${\cal N}=\sum_{{\bm k}, i}1$.)
In the $J=2$ bilayer case, the Hartree potential favors smaller total polarization so that the initial normal (N) state
instability (blue separatrix) is to
antiferromagnetic (AF) states in which the pseudospin polarizations of different valley and spin components cancel.  At larger
$\alpha$, the normal state is unstable (green separatrix) to ferromagnetic (F) pseudospin states.
In the $J=1$ monolayer case ${\bar d}=0$ so the phase boundaries (red separatrix) of F and AF broken-symmetry states coincide.\label{fig:phase}}
\end{center}
\end{figure}

\noindent {\it Pseudospin magnet phase diagram}---
We test the stability of the ``normal" state [$n^{(i)}_z(k) \equiv 0$ at $V_{\rm g}=0$] solution of the Hartree-Fock
equations by linearizing the self-consistency condition; $n^{(i)}_z = B^{(i)}_z n^{(i)}_{\perp}/B^{(i)}_{\perp} \to
B^{(i)}_z / B^{(i)}_{\perp}\vert_{n^{(i)}_z\equiv 0}$.  This gives a $k$-space integral equation
\begin{equation}\label{eq:linearized_equation}
n^{(i)}_z(k) = \sum_j \int_{0}^{1} \; dk'  \; M_{i, j}(k,k') \; n^{(j)}_z(k'),
\end{equation}
where
\begin{equation}\label{eq:linearized_matrix}
M_{i, j}(k,k') =  \frac{\alpha [F_{z}(k,k')\delta_{i, j} - k' {\bar d}/2] \;f^{(j)}_{\rm diff}(k')}{\displaystyle k^{J} +  \alpha \int_{0}^{1} dk'' F_{\perp}(k,k'') \; f^{(i)}_{\rm diff}(k'')}\,.
\end{equation}
The normal state is stable when the largest eigenvalue of the linear integral
operator ${\cal M}$ in the right-hand side of Eq.~(\ref{eq:linearized_equation})
is smaller than $1$.
%$M_{i, j}(k,k')$ can first be diagonalized in the $i,j$ space to take care of the component degree of freedom with the result that in the Hartree term %which opposes pseudospin polarization $k'{\bar d} \to 4 k' {\bar d}$ for the ``ferromagnetic" mode in which the four components are layer polarized in %the same sense, and $k' {\bar d} \to 0$ for the two independent ``antiferromagnetic" modes in which the net pseudospin polarization is zero.}
Eigenvalues larger than $1$ are possible only because $F_{\perp}$ is smaller than $F_z$, {\em i.e.}, because of
pseudospin chirality.  Phase diagrams for $J=2$ and $J=1$ are plotted
in Fig.~\ref{fig:phase}.  The pseudospin magnet is more stable for larger coupling constant because it
is driven by interactions, for larger $J$ because the typical value of the band energy term proportional to $k^{J}$ decreases
with $J$, and for smaller doping because $f^{(i)}_{\rm diff}(k)$ is then nonzero in a larger region of
$k$-space.  The eigenvectors of ${\cal M}$ specify the instability channel.  The component-index structure of ${\cal M}$ implies that
the eigenvalues occur in groups of four, three of which [labeled {\em antiferromagnetic} (AF) in Fig.~\ref{fig:phase}] correspond
in bilayers to states with no net charge transfer, {\em i.e.}, $\sum_j n^{(j)}_z(k) \equiv 0$. The {\em ferromagnetic} (F) instability in which
all components are polarized in the same sense is opposed by the Hartree potential and delayed to larger coupling constant.
For both AF and F instabilities, $n_z(k)$ is peaked at small $k$
where the ${\hat x}-{\hat y}$ pseudospin-plane exchange energies are most strongly
frustrated by chirality, and the kinetic energy term which opposes pseudospin magnetism is weakest.
%The system is initially unstable to states with no net pseudospin polarization, as indicated in
%Fig.~\ref{fig:phase}, but states with full pseudospin-polarization are also metastable and
%expected to appear as gate voltage is swept.

\begin{figure}
\includegraphics[width=1.0\linewidth]{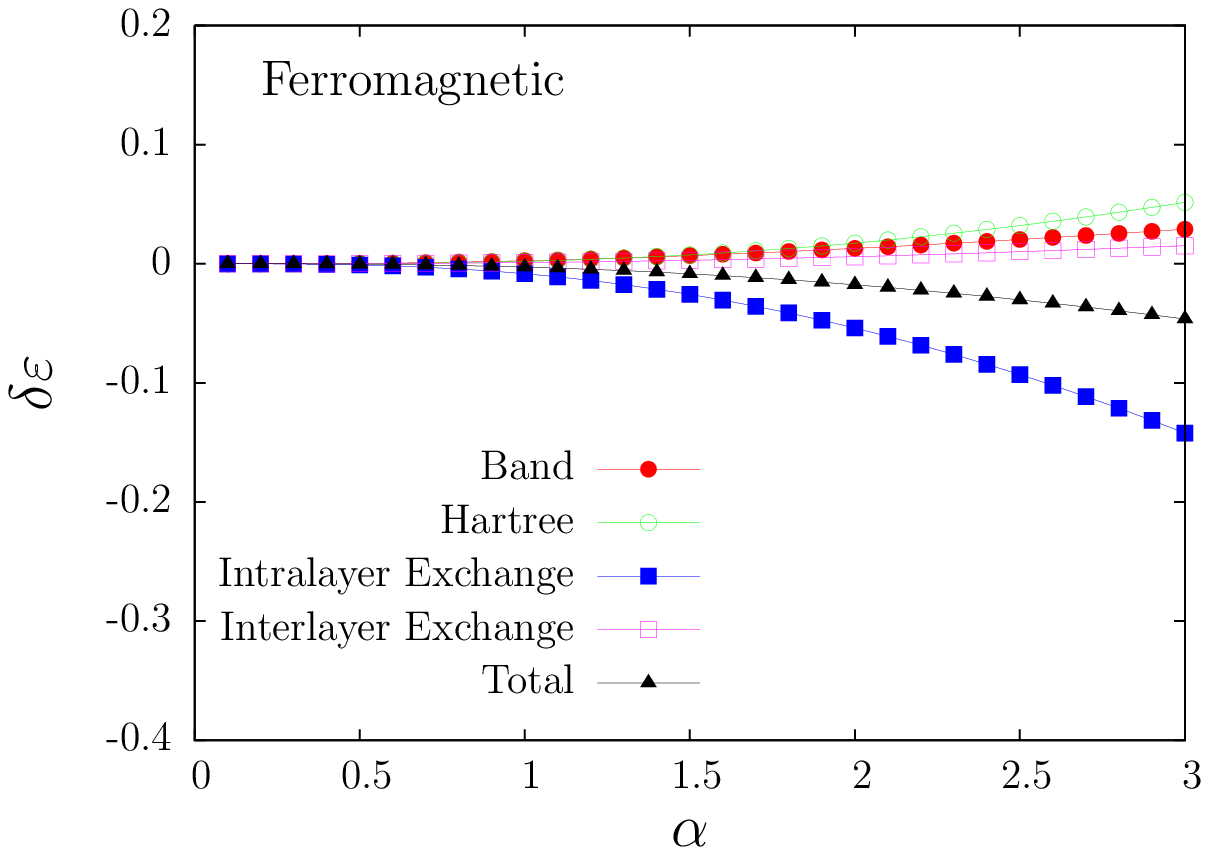}
\includegraphics[width=1.0\linewidth]{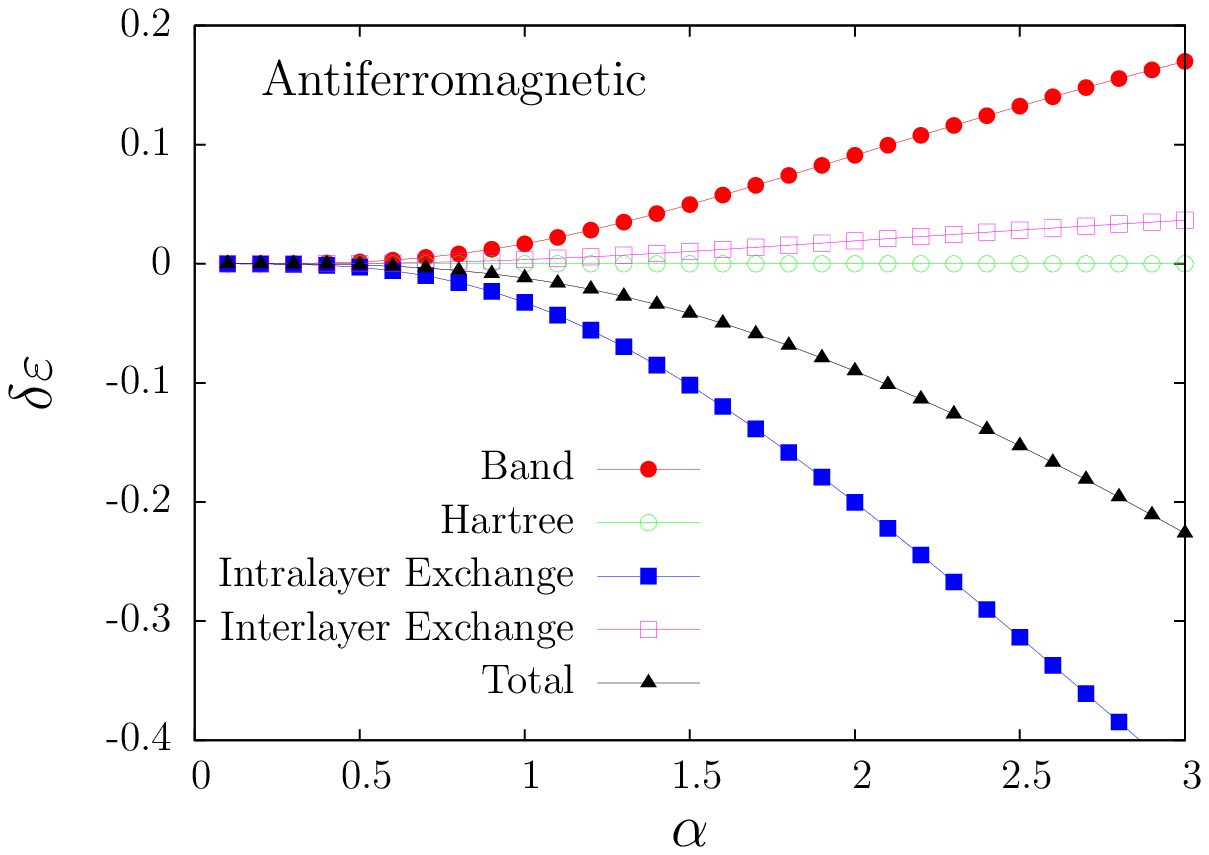}
\caption{(Color online) Condensation energy per electron $\delta \varepsilon$ [in units of $\varepsilon_0(k_{\rm c})$]
as a function of $\alpha$ for an undoped ($f=0$) $J=2$ C2DES.
This figure shows results for both F (top) and AF (bottom) states.
%Symmetry is broken because the gain in intralayer exchange energy due to pseudospin
%polarization exceeds the cost in interlayer exchange energy, Hartree energy, and band energy.  For the
%pseudospin antiferromagnet (bottom panel) there is no Hartree energy penalty,
%and the total condensation energy is larger in absolute value.
\label{fig:energy_partition}}
\end{figure}

The physics that drives pseudospin magnetism in graphene bilayers is illustrated in
Fig.~\ref{fig:energy_partition} which partitions the condensation energy into band,
Hartree, intralayer exchange, and interlayer exchange contributions.
Spontaneous layer polarization lowers the intralayer interaction energy at a cost
in all other components.   The overall energy change is negative, and the broken symmetry state
occurs, because the interlayer exchange energy of the normal state is weakened by
the band-Hamiltonian induced frustration explained earlier.  The cost in interlayer exchange energy
of pseudospin rotation is therefore much smaller than the gain in intralayer exchange
energy and the overall energy is reduced.  The energy gain is considerably larger for AF broken symmetry states.

\begin{figure}
\includegraphics[width=1.0\linewidth] {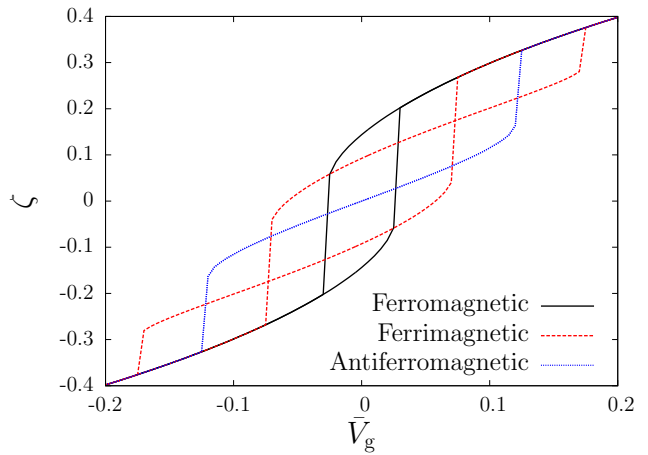}
\caption{(Color online) Metastable configurations of the pseudospin ferromagnet as a function of bias voltage $V_{\rm g}$
[in units of $\varepsilon_0(k_{\rm c})$] with $\alpha=1$ and $f=0$.
We find self-consistent solutions of the gap equations (\ref{eq:Bperp})-(\ref{eq:Bz}) in which the pseudospin polarization has the same sense
in all four components (ferromagnetic), in three of the four components (ferrimagnetic), or in half of the four components (antiferromagnetic).\label{fig:hysteresis}}
\end{figure}

\noindent {\it Pseudospintronics}--- In Fig.~\ref{fig:hysteresis} we illustrate typical results for the
pseudospin (layer) polarization $\zeta=(n_\uparrow-n_\downarrow)/(n_\uparrow+n_\downarrow)$
of a graphene bilayer as a function of gate voltage ${\bar V}_{\rm g}$.
%(the pseudospin densities $n_\sigma$ have been defined in the caption to Fig.~\ref{fig:phase})$.
The AF ground state at ${\bar V}_{\rm g}=0$, which has $[Z_2 \times SU(4)]/[SU(2)\times SU(2)]$ broken symmetry because of the freedom to choose
any two spin or pseudospin components for (say) positive polarization, is gradually polarized by the gate voltage, but eventually
becomes unstable in favor of polarizing more layers in the sense preferred by the gate voltage.  At sufficiently strong
gate voltages, the F ground state in which all layers are polarized in the same sense becomes the ground state.  As the gate voltage
is varied local minima of the Hartree-Fock energy functional become saddle points which are in the basin of attraction of
another local minima.  In this way, the self-consistent solutions exhibit hysteretic behavior.

If only the fully polarized solutions existed these results for pseudospin polarization as a function of gate voltage would be very much
like the behavior expected for an easy-axis ferromagnet in an external magnetic field along the hard axis.
In magnetic memories bistability enables
information storage.  In magnetic metal spintronics the dependence of the resistance of a circuit containing magnetic elements on the
magnetization orientation of those elements gives rise to sudden changes in resistance with field (giant magnetoresistance)
which can be used to sense very small magnetic fields.  Currents running through such a circuit can also be used to change the magnetic
state through spin-transfer torques.  Pseudospin ferromagnetism in graphene bilayers could potentially lead to very
appealing electrical analogs of both of these effects.  Because of the collective behavior of many electrons,
the pseudospin ferromagnet can be switched between metastable states with gate voltages that are much smaller than the
thermal energy $k_{\rm B} T$, potentially enabling electronics which is very similar to standard complementary metal-oxide-semiconductor but uses much less power.
This possibility is analogous to the property that a magnetic element can be switched between magnetic states by
Zeeman field changes that are extremely small compared to the thermal energy $k_{\rm B} T$.  Pseudospin-transfer torques,
which are expected to occur in electronic bilayer systems~\cite{abedinpour2007}, can also be used to switch the
pseudomagnetic state.

% Fig 4:  Bands at finite Bias voltage
%\begin{figure}
%\includegraphics[width=1.0\linewidth] {fig:bands}}
%\caption{Pseudospin bands and polarizations indicated somehow for a bilayer with small
%but nonzero doping.  You can do the fully polarized states.  I am not certain that we'll have room to
%include this plot.}
%\scalebox{0.45}{\includegraphics{band_nz}}
%\caption{(Color online) Pseudospin bands for different degeneracy factors with $g=1$ and $f=0$ at $V_g=0$.}
%\label{fig:band_nz}
%\end{figure}

\noindent {\it Discussion}--- The proposals made here are based on approximate calculations and must
ultimately be confirmed by experiment.  Indeed, it is well known that
Hartree-Fock theory (HFT) often overestimates the tendency toward broken symmetry states.
For example HFT predicts that a non-chiral 2DES
is a (real-spin) ferromagnet at moderate coupling strengths, whereas experiments and accurate quantum Monte Carlo
calculations suggest that ferromagnetism occurs only at a quite large value of the coupling constant~\cite{Giuliani_and_Vignale}.
Nilsson {\em et al.}~\cite{nilsson_2006} have recently
claimed that a similar ferromagnetic instability occurs in weakly-doped graphene bilayers,
presumably only at a much stronger coupling constant than implied by HFT.
We believe that the momentum-space vortex instability identified here, which is
unique to the peculiar band-structure of bilayer graphene, is qualitatively more robust than the real-spin 
ferromagnetic instability.  This should be especially true in neutral bilayers since
the momentum-space vortex instability occurs at a coupling constant ($\alpha \to 0$) for which correlation corrections to HFT are weak.
This is not a strong-coupling instability like ferromagnetism,
but much more akin to the very robust attractive-interaction weak-coupling
instability which 
%for attractive interactions 
leads to superconductivity.
The condensation energy per electron associated with the formation of a
momentum-space vortex core is $\sim e^2 k_{\rm c}/\epsilon$, much larger than the $\sim e^2 k_{\rm F}/\epsilon$ condensation energy for the spin-polarized state.
Because this broken symmetry state is most robust for uniform neutral bilayers, the smooth
but strong disorder potentials responsible for inhomogeneity~\cite{yacoby_2007} in nearly neutral
graphene sheets may need to be limited to allow this physics to emerge.   

\acknowledgements
Work at UT Austin was supported by the Welch Foundation, NRI-SWAN, ARO, and DOE.

\end{document}